# Inverse-current quantum electro-oscillations in a charge-density-wave insulator

Tian Le[1†], Ruiyang Jiang[1,2†], Linfeng Tu[1,3], Renji Bian[4], Yiwen Ma[1,2], Yunteng Shi[1,2], Ke Jia[1,2], Zhilin Li[1], Zhaozheng Lyu[1], Xuewei Cao[3], Jie Shen[1,5], Guangtong Liu[1,5], Youguo Shi[1,2,5], Fucai Liu[4,6], Yi Zhou[1,5,7], Li Lu[1,2,5*], Fanming Qu[1,2,5*]

[1]Beijing National Laboratory for Condensed Matter Physics, Institute of Physics, Chinese Academy of Sciences, Beijing 100190, China

[2]School of Physical Sciences, University of Chinese Academy of Sciences, Beijing 100049, China

[3]School of Physics, Nankai University, Tianjin 300071, China

[4]School of Optoelectronic Science and Engineering, University of Electronic Science and Technology of China, Chengdu 611731, China

[5]Songshan Lake Materials Laboratory, Dongguan, Guangdong 523808, China

[6]Yangtze Delta Region Institute (Huzhou), University of Electronic Science and Technology of China, Huzhou 313009, China

[7]Kavli Institute for Theoretical Sciences & CAS Center for Excellence in Topological Quantum Computation, University of Chinese Academy of Sciences, Beijing 100190, China

[†]These authors contributed equally to this work.

[*]Emails: lilu@iphy.ac.cn; fanmingqu@iphy.ac.cn

**Quantum magneto-oscillations have long been a vital subject in condensed matter physics, with ubiquitous quantum phenomena and diverse underlying physical mechanisms. Here, we demonstrate the intrinsic and reproducible DC-current-driven quantum electro-oscillations with a periodicity in the inverse of the current ($1/I$), in quasi-one-dimensional charge-density-wave (CDW) insulators $(TaSe_4)_2I$ and $TaS_3$ nanowires. Such oscillations manifest in the nearly infinite Fröhlich conductivity region where the undamped CDW flow forms in a finite electric current, and finally disappear after the oscillation index $n$ reaches 1. A systematic investigation on the effect of temperature and magnetic field establishes that the**

**observed electro-oscillations are a coherent quantum phenomenon. We discuss the possibilities of the physical mechanisms, including the formation of sliding-driven inherent Floquet sidebands. Our results introduce a new member in the family of quantum oscillations, and shed light on plausible avenues to explore novel physics and potential applications of coherent density-wave condensates.**

## I. INTRODUCTION

Observables oscillating with quantized indices are closely related to quantum coherence phenomena in physics. For instance, the oscillatory superconducting transition temperature $T_c(B)$ observed in the Little-Parks experiment is a periodic function of the magnetic field $B$, manifesting the macroscopic quantum coherence in a superconductor[1]; and Aharonov–Bohm and Altshuler-Aronov-Spivak oscillations in mesoscopic rings or cylinders arise from quantum interference and have a periodicity in the external magnetic field as well[2,3]. In addition, de Haas-van Alphen (dHvA) and Shubnikov-de Haas (SdH) oscillations arise from Landau quantization and are periodic in the inverse of the magnetic field ($1/B$)[4,5]. Moreover, a variety of possible new quantum magneto-oscillations have recently emerged in different condensed matter systems, such as high-temperature quantum oscillations caused by recurring Bloch states[6-8], log($B$)-period quantum oscillations arising from discrete-scale invariance[9-11], giant quantum oscillations resulting from chiral zero sound[12,13], and extraordinary quantum oscillations in particular insulators caused by quantized orbital motion or charge-neutral fermions[14,15]. In addition to the beauty and the fundamental research interest of these phenomena, quantum magneto-oscillations have also served as powerful techniques to probe essential physical properties, such as Fermi surface (FS) and Berry phase in various quantum materials[16-22]. In this work, we demonstrate inverse-current $1/I$ quantum electro-oscillations in a quasi-one-dimensional (quasi-1D) charge-density-wave (CDW) insulator, phenomenologically, that can come into being with totally different physical origins from the textbook $1/B$ quantum magneto-oscillations.

The CDW is an ordered coherent quantum fluid of electrons with a standing wave pattern. Below the so-called Peierls temperature (CDW transition), the ground state is characterized by a gap in the single-particle excitation spectrum with a collective mode formed by electron-hole pairs[23,24]. In most materials, the CDW is pinned to impurities and requires a finite electric field to drive the condensate to slide collectively, generating a nonlinear voltage-current (*V-I*) curve[25]. More interestingly, Fröhlich argued that such electric current can travel dissipationlessly with an infinite conductivity[26]. The CDW state is widespread in many quasi-1D or -2D materials and permeates much of condensed matter physics such as superconductivity and topological properties[27-32].

(TaSe$_4$)$_2$I is a typical incommensurate quasi-1D CDW material that has been studied for a long time[23,24,33-35]. A CDW transition occurs around $T_{\text{CDW}} = 263$ K while the FS is fully gapped, thus showing an insulating behavior below $T_{\text{CDW}}$[35]. At low temperatures, due to the suppression of thermally activated quasiparticles, it is possible to realize quite a pure coherent sliding mode with the assistance of an electric field. Recently, a Weyl semimetal has displayed in (TaSe$_4$)$_2$I at the room temperature phase[31]. More intriguingly, an axionic CDW phase in the sliding mode of (TaSe$_4$)$_2$I in colinear electric and magnetic fields has been reported[31,36], although it is still debated[37]. Here, we focused on the sliding mode in (TaSe$_4$)$_2$I nanowires, and observed unambiguous quantum electro-oscillations with a periodicity in 1/*I* by current-driven *V-I* measurements, which was further verified by the reproducibility on another well-established CDW material TaS$_3$. Such oscillations appeared only in the nearly infinite Fröhlich conductivity region and disappeared when the oscillation index reached 1, namely, the quantum limit. The phenomenon vividly resembles the 1/*B* quantum magneto-oscillations arising from Landau quantization.

## II. BASIC TRANSPORT PROPERTIES OF THE DEVICES

The (TaSe$_4$)$_2$I single crystals were synthesized via the chemical vapor transport method using I$_2$ as the transporting agent. The raw materials were sealed in a quartz tube under vacuum. The temperature in the reaction zone was set to 530 °C, while the growth zone was kept at 450 °C for 14 days. Finally, the furnace was naturally cooled down to room temperature and needle-like crystals were collected in the growth zone. The TaS$_3$ single crystals were grown by a similar method. With a quasi-1D van der Waals structure, (TaSe$_4$)$_2$I and TaS$_3$ whiskers are easily exfoliated down to nanowires with a width ($d$) of tens to hundreds of nm. Ti/NbTiN (5 nm/100 nm) or Ti/Nb (5 nm/100 nm) electrodes were deposited by magnetron sputtering followed by soft plasma cleaning, using standard electron-beam lithography techniques. For the characterization of bulk whiskers, silver paint was used for the contacts. The resistance, narrow band noise (NBN), and *V-I* curve measurements of the devices were performed in an Oxford TeslatronPT system equipped with a 14 T superconducting magnet. π filters consisting of two 1 nF capacitors and a 10 mH inductor were mounted for all the DC electrical lines at room temperature, when used. In the following, we will focus on the results of (TaSe$_4$)$_2$I bulk whiskers and nanowire devices D1-D4, and present results of TaS$_3$ briefly to demonstrate reproducibility.

NBN is a typical characteristic of the coherent sliding CDW mode[38], as has been measured in both (TaSe$_4$)$_2$I and TaS$_3$ bulk whiskers[35,39]. The fundamental frequency *f* generally exhibits a linear correlation with the CDW current density. To characterize our materials, we performed the NBN measurements on (TaSe$_4$)$_2$I and TaS$_3$ bulk whiskers, as shown in Fig. 1. Figures 1(a) and 1(c) show typical NBN curves under different driven current $I_{\text{total}}$, where the fundamental noise peak could be clearly recognized. Figures 1(b) and 1(d) display the nearly linear dependence of the frequency *f* on the CDW current $I_{\text{CDW}}$ (see Appendix B for the extraction of $I_{\text{CDW}}$). These results reproduce the main features of NBN in the early literatures well[35,39] and verify the quality of our materials.

Another characteristic of CDW in (TaSe$_4$)$_2$I or TaS$_3$ is the insulating behavior below the transition temperature $T_{\text{CDW}}$. We have characterized a great number of (TaSe$_4$)$_2$I nanowire devices and all of them showed a CDW transition around $T_{\text{CDW}} = 263$ K

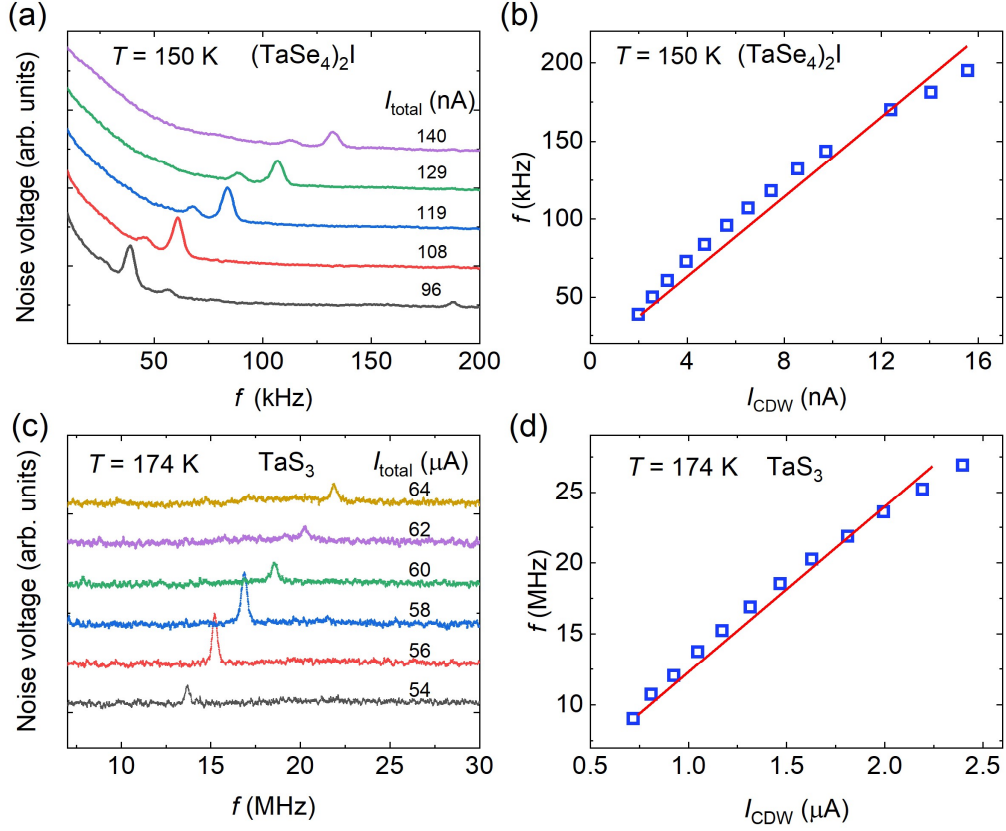

FIG. 1. (a, c) Typical NBN curves of a $(TaSe_4)_2I$ bulk whisker and a $TaS_3$ bulk whisker measured at $T = 150$ K and 174 K, respectively. (b, d) Dependence of NBN frequency $f$ on CDW current $I_{CDW}$. The red lines are a guide to the eye.

(see Appendix B for $TaS_3$). The resistance ($R$) measurement for $(TaSe_4)_2I$ was conducted in a four-probe configuration by applying a constant DC current on the sample and measuring the voltage. Figure 2(a) shows a representative temperature-dependent resistance curve for a 340 nm wide nanowire, which has been plotted as logarithmic normalized resistance versus $1/T$. Note that all electrodes used in our $(TaSe_4)_2I$ nanowire devices in the main text are Ti/NbTiN with a superconducting transition temperature $T_c \sim 12$ K and an upper critical field $H_{c2} > 13$ T at 1.55 K. The choice of the sputtered Ti/NbTiN was aimed at improving the contact quality, as discussed in the Appendix A. Below $T_{CDW}$, the increase of the resistance originates from FS gapping by the CDW and initially follows the thermal activation law, $\frac{R}{R_{300\,K}} \approx e^{\Delta/k_B T}$ for single-particle state, where $k_B$ is the Boltzmann constant[35,36]. The single-particle gap $\Delta \sim 165$ meV can be extracted from a linear fit of the experimental data as shown in Fig. 2(a), where the deviation from the activation law at low temperatures is presumably due to the participation of the sliding CDW mode at a fixed driven current of 0.1 μA.

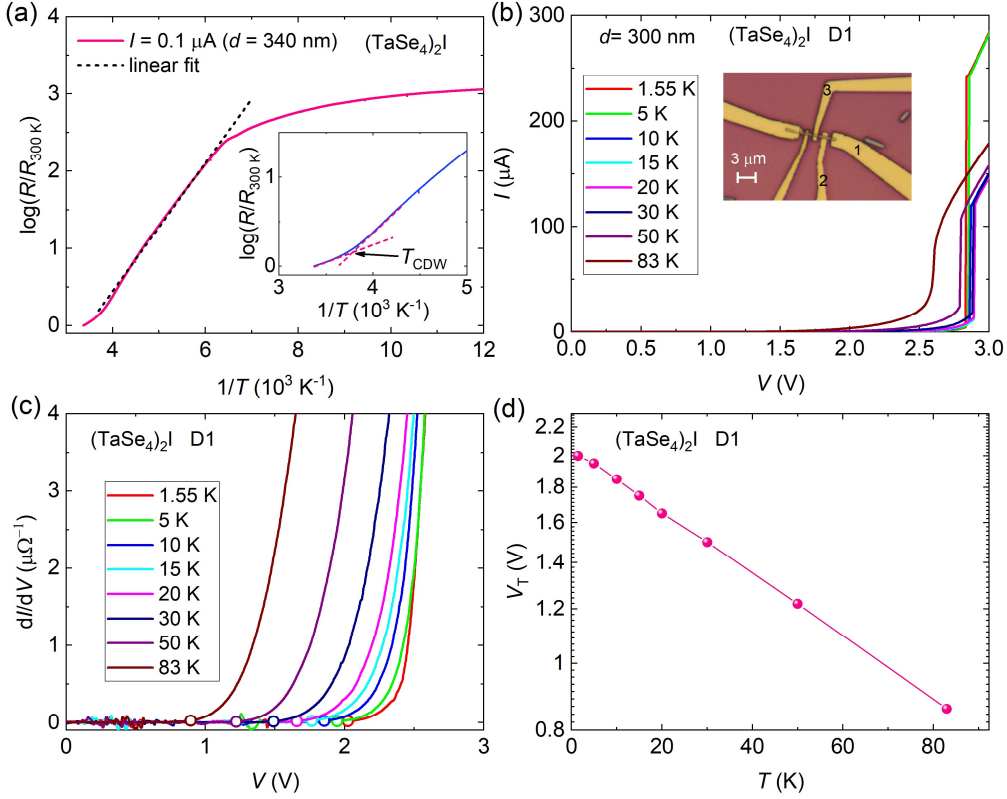

FIG. 2. (a) Logarithm of normalized resistance as a function of $1/T$ for a typical $(TaSe_4)_2I$ nanowire with a width $d = 340$ nm, measured with a DC current of 0.1 μA. The dashed line is a linear fit of the experimental data. The inset presents a CDW transition around $T_{CDW} = 263$ K. (b) Voltage-driven $I$-$V$ curves measured between electrodes 1 and 3 on device D1 at different temperatures. The inset shows the optical microscope image of D1. (c) Differential conductance $dI/dV$ vs. $V$ curves extracted from (b). The circles indicate the threshold voltage $V_T$. Note that for clarity only the bottom part of each curve is shown. (d) The $V_T$ vs. $T$ curve, which shows an exponential dependence.

To study the threshold voltage $V_T$ for the depinning of the CDW, we measured $I$-$V$ curves on $(TaSe_4)_2I$ nanowire device D1 at different temperatures in a voltage-driven mode, as shown in Fig. 2(b). The current $I$ is nearly zero when $V$ is small, and starts to increase at a threshold $V_T$. When $V$ increases further, there is a sudden jump (switch), indicating the Fröhlich conductivity regime[26,40-44], as explained later. To extract the threshold $V_T$, we plot the differential conductance d$I$/d$V$ vs. $V$ curves in Fig. 2(c), where the circles indicate $V_T$. We observed the well-established exponential dependence of $V_T$ on temperature $T$[23,24,35,38], as shown in Fig. 2(d).

All these characterizations verify the quality of our materials and devices.

## III. 1/$I$-PERIODIC OSCILLATIONS

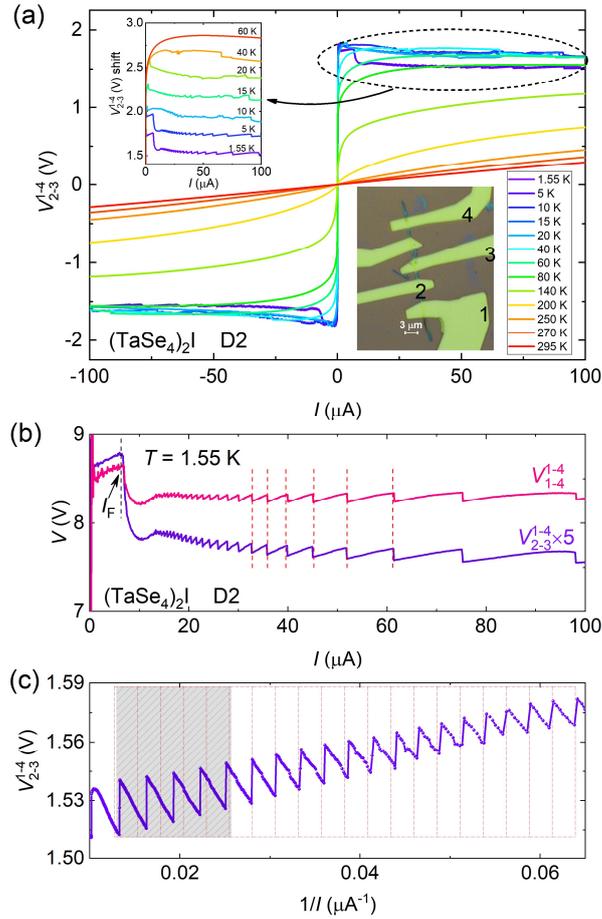

FIG. 3. (a) Temperature dependence of the $V_{2-3}^{1-4}$-$I$ curves for (TaSe$_4$)$_2$I nanowire device D2 whose optical image is shown in the inset with a nanowire width of ~300 nm. (b) $I$ dependence of $V_{2-3}^{1-4}$ and $V_{1-4}^{1-4}$ at 1.55 K. $I_F$ represents the threshold current signifying the onset of Fröhlich conductivity, as delineated by the back arrow. (c) $V_{2-3}^{1-4}$ as a function of 1/$I$, exhibiting periodic oscillations/jumps. The shaded area indicates the deviations from the periodic oscillations at high current.

Next, we examine the DC-current-driven *V-I* curves from room temperature to low temperature on (TaSe$_4$)$_2$I nanowire device D2. In the main text, $V_{b-c}^{a-d}$ denotes the voltage measured between electrodes b and c, while the current is applied between electrodes a and d. As shown in Fig. 3(a), when the temperature crosses $T_{CDW}$, *V-I* curves change from being linear to nonlinear, signaling the sliding of CDW when the voltage $V_{2-3}^{1-4}$ exceeds a threshold voltage $V_T$. At lower temperatures, as shown in the inset of Fig. 3(a), the *V-I* curves become nearly flat at large currents, indicating a nearly infinite differential conductivity which could be attributed to the Fröhlich mechanism of CDW condensates in a coherent sliding mode[26,40-44]. In particular, sudden voltage drops in the Fröhlich conductivity region gradually appear as the temperature is further decreased, which finally behave as quite regular voltage oscillations/jumps at 1.55 K, as explicitly shown in Fig. 3(b). When the current was applied between electrodes 1 and 4, the four-probe ($V_{2-3}^{1-4}$) and two-probe ($V_{1-4}^{1-4}$) measurements showed coincident oscillations/jumps, as marked by the dashed lines in Fig. 3(b). Note that the accurate measurement of voltage at ultralow current is hindered by the giant resistance when the CDW is pinned. Hence, extracting $V_T$ merely from current-driven *V-I* curves in Fig. 3 is challenging. Instead, we have verified the exponential relationship of $V_T$ on *T* by voltage-driven measurements, as shown in Fig. 2(d). While this $V_T$ signifies the depinning of the CDW, in the current-driven mode, there is a threshold current $I_F$ that implies the initiation of Fröhlich conductivity[44], as denoted by the black arrow in Fig. 3(b). Below $I_F$, CDW has been depinned when the voltage exceeds $V_T$. Above $I_F$, a narrow region of negative differential resistance (NDR) was observed, as has been demonstrated in various CDW materials[38,42,45,46]. Above the NDR region, regular voltage oscillations/jumps were observed in the nearly flat part. Most surprisingly, as depicted in Fig. 3(c), the oscillations/jumps are periodic in 1/*I*, except for the deviation at high currents where a heating effect may play a role (Appendix D). We would like to emphasize that this is a new phenomenon and, to the best of our knowledge, has not yet been reported.

# IV. TEMPERATURE AND MAGNETIC FIELD DEPENDENCES OF THE 1/$I$ OSCILLATIONS

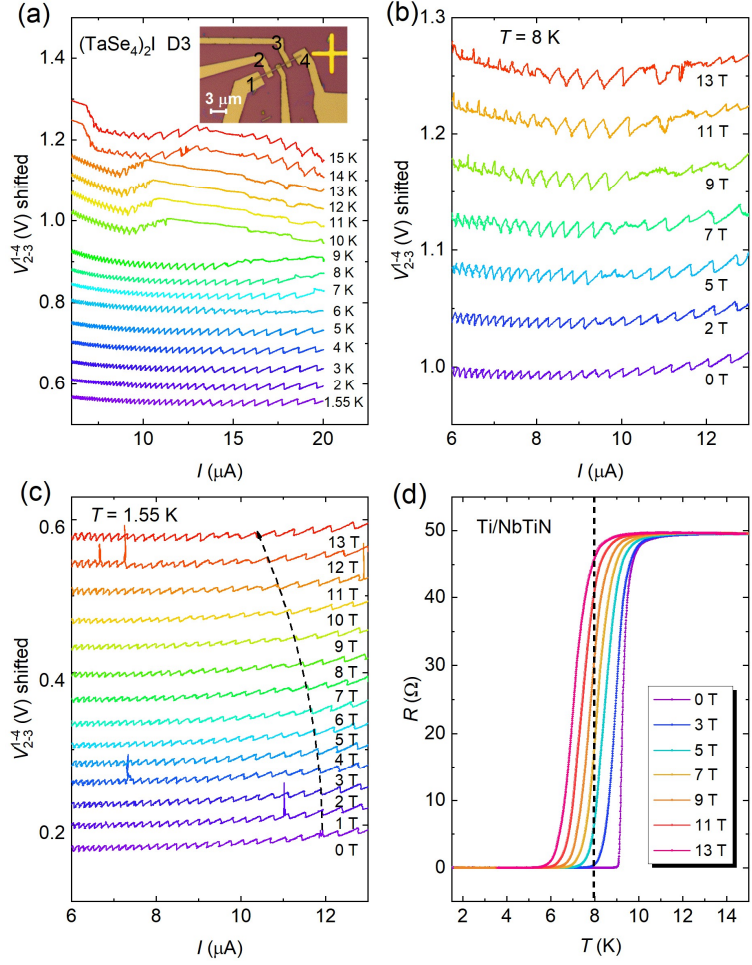

FIG. 4. (a) Temperature dependence of $V_{2-3}^{1-4} - I$ curves for $(TaSe_4)_2I$ nanowire device D3 from 1.55 K to 15 K. The inset is the optical image of D3 with the nanowire width of ~570 nm. (b) Magnetic field evolution of $V_{2-3}^{1-4} - I$ curves from 0 T to 13 T at 8 K. (c) Magnetic field dependent $V_{2-3}^{1-4} - I$ curves from 0 T to 13 T at 1.55 K. Note that there is a slight difference between the bottom curves in (a) and (c) after a thermal cycle. (d) $R$-$T$ curves for a Ti/NbTiN electrode under different magnetic fields. $B$ was applied perpendicular to the substrate in (b) and (d) to kill the superconductivity easily, but was parallel to the substrate in (c) to maintain the superconductivity as far as possible.

We further inspect the temperature and magnetic field dependence of the 1/$I$ oscillations in (TaSe$_4$)$_2$I nanowire device D3, which presents oscillations at lower currents than D2. In Fig. 4(a), the temperature dependent *V-I* curves from 1.55 K to 15 K suggest a gradual smearing behavior with increasing *T*, and a strong smearing above 9 K. Since the zero-resistance state of the Ti/NbTiN electrode persists up to 9 K as shown in Fig. 4(d), we conjecture that the suddenly enhanced dissipation in the electrodes and/or the interfaces may contribute to this smearing except for increasing *T*. Another complementary feature comes from the evolution of the *V-I* curves in magnetic fields at 8 K, as displayed in Fig. 4(b), which shows a smearing behavior of the high-quality 1/*I* oscillations with increasing *B* that destroys the superconducting state of NbTiN. Accordingly, the resistance of the Ti/NbTiN film increases from 3 T to 13 T at 8 K, as measured separately and shown in Fig. 4(d). To eliminate the possible influence of the electrodes, the examination of the magnetic field dependence of the 1/*I* oscillations was further performed at 1.55 K when NbTiN remained superconducting. As depicted in Fig. 4(c), the 1/*I* oscillations are nearly independent on *B*, except for a slight decrease of the current for a certain voltage jump as indicated by the dashed arrow (as discussed in Appendix E). In addition, for other pairs of electrodes in D2, $V_{2-3}^{2-3} - I$, $V_{1-3}^{1-3} - I$ and $V_{2-4}^{2-4} - I$ did not show regular 1/*I* oscillations (Appendix Fig. 12). In accordance with the above observations, it is reasonable to speculate that the 1/*I* oscillations are associated with the synergistic effect of the current contacts, the low temperature, and the CDW. We note that superconducting electrodes are preferred but not indispensable as discussed in Appendix D, where the 1/*I* oscillations were observed in both the superconducting and normal states of Ti/Nb electrodes.

## V. ANALYSIS OF THE OSCILLATION INDEX *N*

Next, we introduce the oscillation index $n$ (an integer). Phenomenologically, periodic 1/*I* oscillations require $I(n + b) = const.$, when compared with the formula $B(n + b) = const.$ for dHvA and SdH oscillations[47], where $|b| < 1$ (Appendix G). Let us look at the oscillations/jumps of device D3 as shown in Fig. 5(a). (The analysis of the oscillations at high temperatures is shown in Appendix H.) By defining $I_n$ as the current corresponding to each voltage jump, $1/I_n - 1/I_{n-1} = \Delta(1/I) = 0.00269\ \mu A^{-1}$. We can obtain the index $n$ for each $I_n$, as indicated from 55 to 19 in Fig. 5(a). Similar to the Landau fan diagram analysis, $1/I_n$ is plotted as a function of $n$ in Fig. 5(b), which could be remarkably fitted by a linear function $1/I_n = 0.00269(n - 0.02)$.

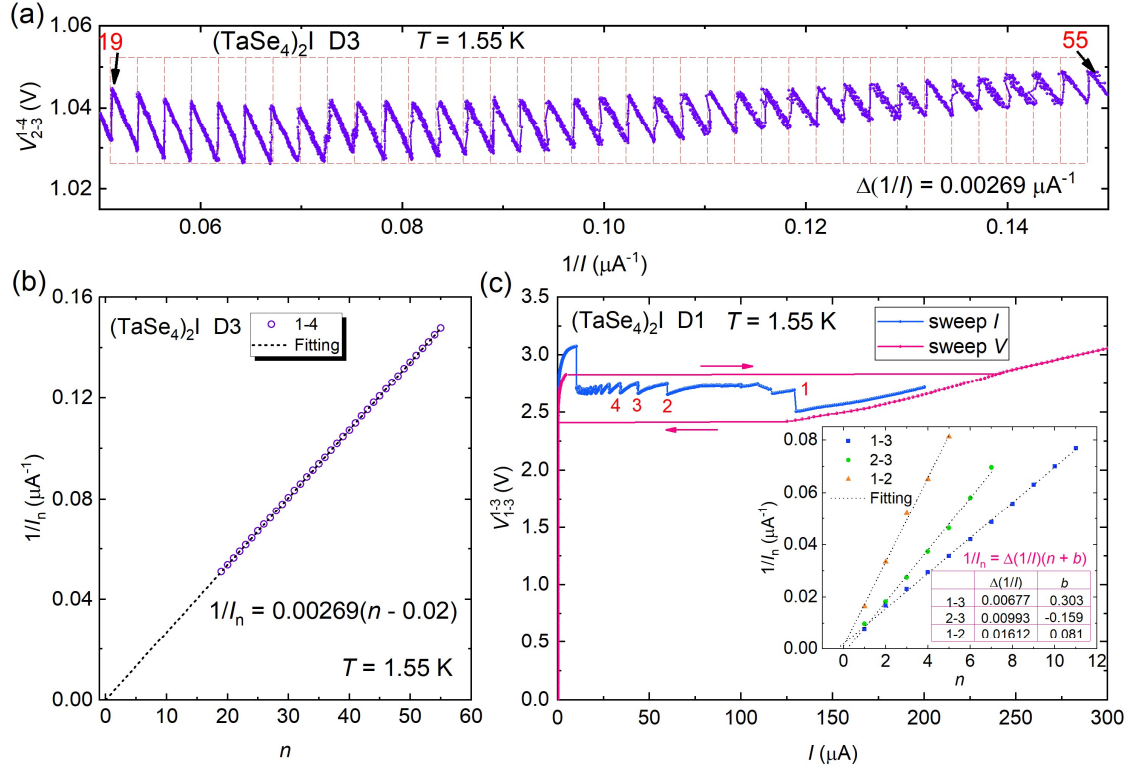

FIG. 5. (a) $V_{2-3}^{1-4}$ vs. $1/I$ for $(TaSe_4)_2I$ nanowire device D3 at 1.55 K with a period of $\Delta(1/I) \sim 0.00269\ \mu A^{-1}$. (b) $1/I_n$ plotted against oscillation index $n$. (c) Current- and voltage-driven V-I curves at 1.55 K for D1. The inset shows $1/I_n$ vs. $n$ for $V_{1-3}^{1-3} - I$, $V_{2-3}^{2-3} - I$ and $V_{1-2}^{1-2} - I$. Black dashed lines in (b) and (c) represent a linear fit to the data.

It is quite intriguing to examine what happens after the index $n$ reaches 1 at high current. However, since the oscillations for device D3 were influenced by strong smearing effect at high current, we examined device D1 [inset of Fig. 2(b)] utilizing three pairs of adjacent electrodes. The index $n$ could be extracted, and again, $1/I_n$ unambiguously shows a linear dependence on $n$ [inset of Fig. 5(c)]. With increasing current, the indices could all reach 1 [Figs. 5(c) and 13]. In particular, as shown by the $V_{1-3}^{1-3} - I$ curve [Fig. 5(c)], the oscillations disappear for $I > I_{n=1}$ and the voltage starts to increase quasi-linearly, vividly mimicking the quantum limit in dHvA and SdH oscillations at high magnetic field[47]. Moreover, the DC-voltage-driven $V_{1-3}^{1-3} - I$ curves display a striking switching behavior, characterized by an abrupt change (increasing or decreasing) of the current and a hysteresis as shown in Fig. 5c. Such switching and hysteretic behavior is also a typical characteristic of CDW[38,48-54]. The current-driven $1/I$ oscillations emerge in the hysteretic region of the voltage-driven $V-I$ curves.

## VI. FILTER EFFECT ON THE 1/$I$ OSCILLATIONS

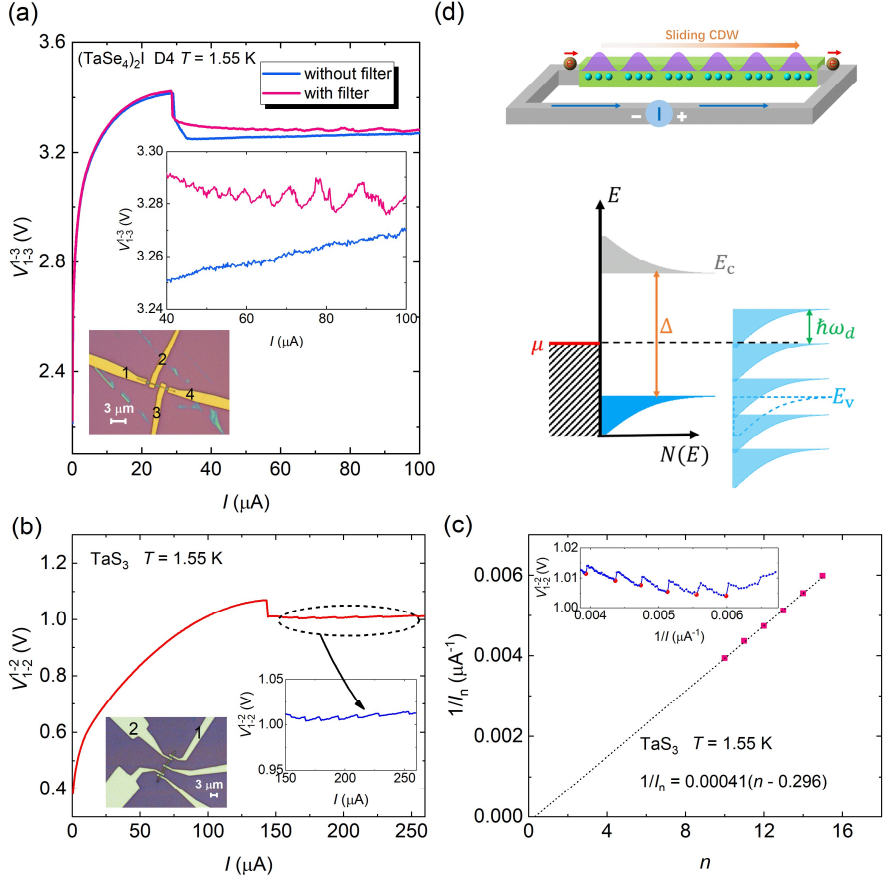

FIG. 6. (a) $V-I$ curves with and without filters in the measurement circuit on $(TaSe_4)_2I$ nanowire device D4 at 1.55 K. Bottom left inset shows the optical image for D4 with a nanowire width of ~350 nm. (b) The left inset shows the device fabricated on an exfoliated $TaS_3$ nanowire using sputtered Ti/Nb electrodes. The main figure presents

the $V_{1-2}^{1-2} - I$ curve measured at 1.55 K. The right inset is a zoom-in of the oscillations/jumps. (c) The inset shows $V_{1-2}^{1-2}$ vs. $1/I$ extracted from (b), i.e., the periodic oscillations/jumps. The main plot displays $1/I_n$ vs. $n$, showing a linear dependence. (d) Top image is a schematic illustration of the sliding-condensate-assisted resonant tunneling. Bottom picture shows the intrinsic Floquet sidebands, which enhances the conductivity when the band-edge aligns with the chemical potential of the incident electrons.

We also present the peculiar effect of filters on the observability of the $1/I$ oscillations. It turns out that the explicit oscillations can be conveniently observed when the measurement circuits are equipped with filters that reduce the electrical noise. Figure 6(a) illustrates a comparison of filtered and unfiltered *V-I* curves of device D4 at 1.55 K. In general, the filters are beneficial for observing high-quality $1/I$ oscillations. This suggests that the $1/I$ oscillations are closely related to the coherence of the CDW condensates between the current electrodes, which is extremely sensitive to external electrical perturbations, similar as the quantum metallic states in superconducting thin films[55]. We must stress that the $1/I$ oscillations could be observed without filters as shown in Appendix Fig. 17 (the possible artifacts from circuits are excluded in Appendix I).

## VII. REPRODUCIBILITY OF THE 1/*I* OSCILLATIONS

We emphasize that the $1/I$ oscillations are an intrinsic phenomenon and can be reproduced on a large number of devices. In total, we have measured 78 $(TaSe_4)_2I$ nanowire devices, among which at least 13 devices exhibited regular periodic $1/I$ oscillations, 11 devices showed irregular oscillations similar to $V_{2-3}^{2-3} - I$, $V_{1-3}^{1-3} - I$ and $V_{2-4}^{2-4} - I$ on D2, as shown in Appendix Fig. 12, and the others were damaged before performing systematic measurements, e.g., when sweeping to only a few µA. To verify the universality of such oscillations, we also measured similar devices on $TaS_3$ nanowires, in which the $1/I$ oscillations were reproduced, as shown in Figs. 6(b) and 6(c). The left inset of Fig. 6(b) shows the device on a $TaS_3$ nanowire using sputtered

Ti/Nb electrodes. The main plot of Fig. 6(b) presents the $V_{1-2}^{1-2}$ – $I$ curve measured at 1.55 K, and a zoom-in of the oscillations/jumps is shown in the right inset. Figure 6(c) plots $V_{1-2}^{1-2}$ vs. 1/$I$, and $1/I_n$ vs. $n$, showing the 1/$I$ periodic oscillations. In general, the appearance of the periodic 1/$I$ oscillations in mesoscopic (TaSe$_4$)$_2$I and TaS$_3$ nanowires is related to contact conditions, temperatures far below $T_{CDW}$, and the coherence of the sliding CDW modes. These peculiarities constrain the theoretical models to explain the underlying mechanisms.

## VIII. DISCUSSION

Now we discuss the possible physical mechanisms for the 1/$I$ oscillations. One can immediately think of NBN and the associated synchronization with an external alternating field. The sliding CDW mode has been verified through the NBN measurements[38,39]. In this context, when subjected to an external alternating field of frequency $f_{ac}$, current-driven oscillations can occur due to mode locking[56-58]. However, these oscillations require a finite power of the external field, and exhibits d$V$/d$I$ peaks when $f/f_{ac} = p/q$ is satisfied [56,58]. Since in our experiment there is no intentionally applied single-frequency radiation, the oscillations are 1/$I$ periodic, and the oscillations/jumps correspond to negative d$V$/d$I$ (the voltage drops), we can safely rule out such a scenario.

In light of the above considerations, we propose a possible mechanism based on sliding CDW modes, which naturally induces coherent resonant tunneling and sliding-driven inherent Floquet sidebands, resembling the photon-assisted quantum tunneling that has been well studied in the context of quantum optics[59] and semiconductor nanostructures[60]. The idea is that for a sliding CDW the phase of the CDW order parameter $\Delta e^{i\phi}$ becomes time-dependent, i.e., $\phi(t) = \phi_0 - \omega_d t$, where $\phi_0$ does not depend on time, and the frequency $\omega_d$ is related to the current density per chain by $J_{CDW} = -\frac{e}{\pi}\frac{d\phi}{dt} = \frac{e}{\pi}\omega_d$[23]. Thus, as illustrated schematically in Fig. 6(d), a time-periodic potential is experienced by an electron in the background of such a sliding CDW condensate, giving rise to Floquet sidebands for quasiparticles[61] (see also the theoretical model in Appendix J). The quasi-energy for quasiparticles is determined as $E_k = E_F + \left(m + \frac{1}{2}\right)\hbar\omega_d \pm \sqrt{\left(\left[\hbar v_F(k - k_F) + \frac{\hbar\omega_d}{2}\right]\right)^2 + \Delta^2}$, up to half-odd-integer multiples of $\hbar\omega_d$. Note that the time-periodic potential is a result of the coherent Fröhlich flow, similar to the voltage-driven Josephson effect, rather than the external radiation as used in the context of Floquet engineering[62]. Whenever a sideband of the gap edge with a singular density of states aligns with the chemical potential $\mu$ of the incident electrons, an enhancement of the conductivity and thus a voltage drop is expected, generating oscillations/jumps periodically in $1/\omega_d$ which is exactly proportional to 1/$I$.

Assuming such a sliding CDW picture, the periodicity $\Delta(1/I)$ depends on the contact-related and self-adapted mismatch $\varepsilon$ between the chemical potential $\mu$ and the lower gap edge $E_v$ [see Fig. 6(d)], i.e., $\varepsilon = \mu - E_v$. Because of the singular density of states at the edge of the CDW gap, the jumps occur at $\varepsilon = \left(n + \frac{1}{2}\right)\hbar\omega_d = \left(n + \frac{1}{2}\right)\hbar\frac{\pi I}{Ne}$, and thus for two adjacent jumps at $I_n$ and $I_{n-1}$, $\varepsilon = \left(n + \frac{1}{2}\right)\hbar\frac{\pi I_n}{Ne} = \left(n - 1 + \frac{1}{2}\right)\hbar\frac{\pi I_{n-1}}{Ne}$, where $N$ is the number of conduction electron channels, giving the periodicity $\Delta\left(\frac{1}{I}\right) \equiv \frac{1}{I_n} - \frac{1}{I_{n-1}} = \frac{\hbar\pi}{Ne\varepsilon}$. Note that the mismatch $\varepsilon$ and the number $N$ can vary in different pairs of electrodes. This explains the different periods on a given nanowire, e.g., the inset of Fig.5(c).

Any decoherence effects are detrimental for the resonant-tunneling and will destroy the observed inverse-current quantum electro-oscillations. Some decoherence mechanisms are discussed below: (1) Thermal fluctuations at higher temperatures are always fatal to phase coherence. (2) The contact of the electrodes/interfaces is another issue. (3) Electromagnetic noise from the environment is also critical and can be eliminated by filters. (4) For a two or three-dimensional CDW sample, there is always a residual FS due to the non-perfect FS nesting, so that they exhibit metallic behavior at low temperatures. In this case, the decoherence effect is due to the scattering between the quasiparticles around the gapped FS and the electrons around the remaining FS. Therefore, a full CDW gap and a coherent Fröhlich flow are crucial for the observation of the quantum-electro oscillations. It is worth noting that all these decoherence mechanisms are suppressed in the $(TaSe_4)_2I$ and $TaS_3$ nanowires used in our experiments. In particular, both $(TaSe_4)_2I$ and $TaS_3$ are quasi-one-dimensional CDW insulators.

Finally, we would like to emphasize that our model captures the $1/I$ periodicity, but cannot explain all the observations, such as the re-entrance of the oscillations at high temperatures (see Appendix H). A well-developed theoretical model is still highly in demand.

## IX. CONCLUSION

In conclusion, we have observed a new type of intrinsic quantum electro-oscillations, periodic in $1/I$ in a mesoscopic quasi-1D CDW system. Such exotic quantum electro-oscillations are preliminarily ascribed to resonant-tunneling associated with intrinsic Floquet sidebands in the Fröhlich conductivity region. Our results reveal a new quantum phenomenon hidden in insulating density-wave materials. It provides a fresh perspective to study the novel physics in coherent condensates, e.g., CDW and spin-density-wave. And the result may also be helpful to develop potential multifunctional quantum technologies.

## ACKNOWLEDGEMENTS

We would like to thank Xi Dai, Nanlin Wang, Hua Jiang, Rui Yu and Yupeng Li for fruitful discussions. This work was supported by the National Basic Research Program of China through MOST Grants Nos. 2022YFA1403403, 2021YFE0194200, and 2020YFA0309200; by the National Natural Science Foundation of China through Grants Nos. 12074417, 92065203, 92365207, 11874406, 12104489, 11774405, 12034004, and 11527806; by the Strategic Priority Research Program of Chinese

Academy of Sciences, Grants Nos. XDB28000000, XDB33010300; by the K. C. Wong Education Foundation (Grant No. GJTD-2020-01); by the Synergetic Extreme Condition User Facility sponsored by the National Development and Reform Commission; and by the Innovation Program for Quantum Science and Technology (Grant No. 2021ZD0302600).

# APPENDIX

## APPENDIX A: Choice of the electrode material

To fabricate the devices on $(TaSe_4)_2I$ nanowires, we tried different electrode materials. Initially, we used Ti/Au electrodes deposited by electron-beam evaporation. However, it usually gave a very large contact resistance (above 1 MΩ) at room temperature although an *in situ* interface cleaning was applied and the voltage-current (*V-I*) curves showed a non-ohmic behavior, even though $(TaSe_4)_2I$ is a semimetal above the charge-density-wave (CDW) transition temperature $T_{CDW} \approx$ 263 K. At low temperatures, the resistance between two electrodes was too large to measure due to the poor contacts. However, we figured out that the Ti/NbTiN or Ti/Nb electrodes deposited by magnetron sputtering provided ohmic contacts with a contact resistance around several kΩ at room temperature. Noteworthy, an *in situ* soft plasma cleaning was also performed prior to the sputtering to improve the quality of the contact. Moreover, superconducting electrodes usually produce less dissipation in the superconducting state, which could be beneficial for our studies as mentioned in the main text. Therefore, in our $(TaSe_4)_2I$ nanowire devices, all electrodes were fabricated using Ti/NbTiN (~5 nm/100 nm), except for devices D6 and D7 which were constituted by Ti/Nb (~5 nm/100 nm). In addition, we would like to point out that the Josephson effect between two superconducting electrodes in our devices was improbable, due to the insulating behavior of $(TaSe_4)_2I$ at low temperatures and the large separation between the electrodes, and as also reflected by the huge resistance around zero current bias. Ti/Nb was used for $TaS_3$ nanowire devices, and similar analysis applies. For the measurements on bulk whiskers, silver paint was dropped on the whiskers to form good contact.

**APPENDIX B: Basic properties of $(TaSe_4)_2I$ and $TaS_3$ in narrow band noise measurement**

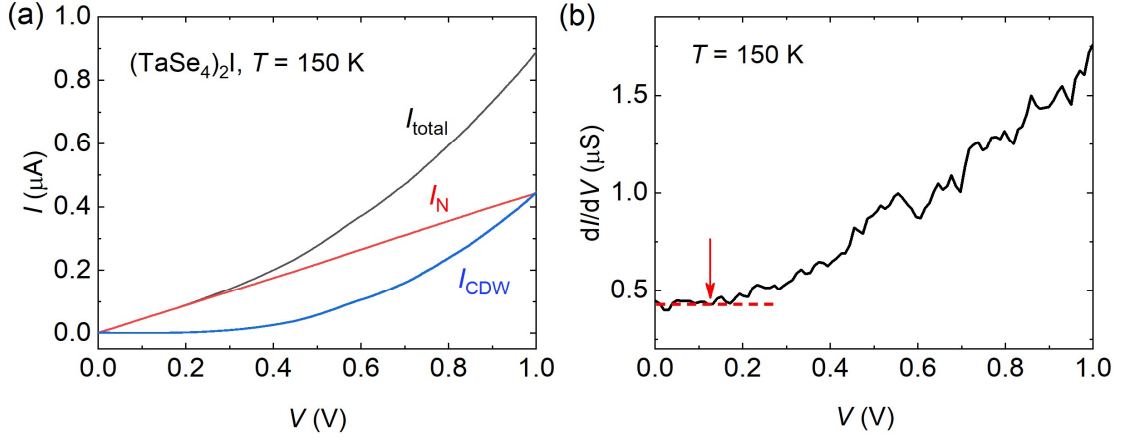

FIG. 7. (a) The $I$-$V$ curve (black) of a $(TaSe_4)_2I$ bulk whisker at $T = 150$ K. The red line shows the current $I_N$ carried by normal carriers, and the blue curve is the current carried by the CDW condensate, $I_{CDW} = I_{total} - I_N$. (b) The $dI/dV$ vs. $V$ curve, where the red dashed line and arrow indicate the normal transport before the threshold voltage. The slope of the red line ($I_N$) in (a) takes the value of this red dashed line.

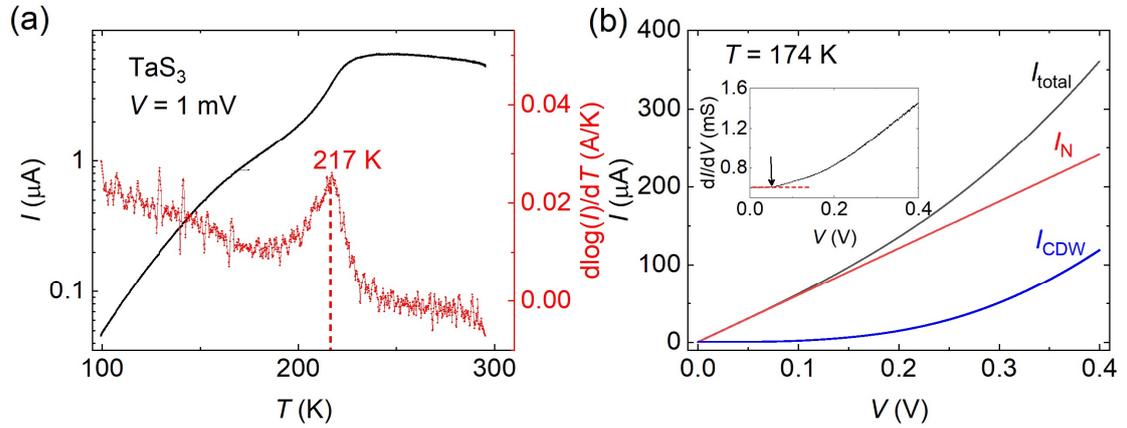

FIG. 8. (a) Temperature dependence of the current $I$ of a $TaS_3$ bulk whisker measured by applying a voltage $V = 1$ mV. The red curve shows $d\log(I)/d(T)$ vs. $T$, where a CDW transition at 217 K can be recognized. (b) The $I$-$V$ curve (black) of the $TaS_3$ bulk whisker at $T = 174$ K. The red line shows the current $I_N$ carried by normal carriers, and the blue curve is the current carried by the CDW condensate, $I_{CDW} = I_{total} - I_N$. The inset shows the $dI/dV$ vs. $V$ curve, where the red dashed line and arrow indicate the normal transport before the threshold voltage. The slope of the red line ($I_N$) takes the value of this red

dashed line.

**APPENDIX C: The threshold voltage $V_T$**

In the main text, we showed the exponential dependence of the threshold voltage $V_T$ on temperature $T$ in Fig. 2(d). If we consider the length between the electrodes (edge to edge) of ~ 2.4 um and $V_T \approx 2$ V at 1.55 K, the threshold electric field is $E_T = 8.3$ kV/cm. Intuitively, $E_T$(1.55 K) is about three orders larger than bulk (TaSe$_4$)$_2$I at high temperatures (135 K)[39], and is nearly two orders larger than mesoscopic (TaSe$_4$)$_2$I at similarly low temperatures[63]. However, we will show that the large $E_T$ in our experiment is reasonable. There are at least two factors that drastically change the threshold $E_T$. One is temperature. The exponential dependence of $E_T$ on temperature has been well established, and was also observed in our devices (Fig. 2(d)). At low temperatures similar to ours, say ~ 1.5 K, $E_T$ of ~ 100 V/cm was extracted by Cohn et al. for mesoscopic devices[63]. Another is the dimensions of the device. Finite-size effect, different pinning regimes, and crossover between different dimensions have been well established in CDW materials[38,64,65]. For example, Slot et al. showed an enhancement of $E_T$ by more than two orders of magnitude when reducing the cross section from ~ 2 μm$^2$ to 0.002 μm$^2$. For device D1 in our experiment, the width is ~ 0.3 μm, and the thickness is < 0.1 μm, i.e., a cross section < 0.03 μm$^2$. If we take $E_T$ ~ 100 V/cm for (TaSe$_4$)$_2$I with a cross section ~ 8 μm$^2$ [63] as a reference, it is reasonable that $E_T$ increases to 8.3 kV/cm in our case. Recently, $E_T$ as large as 20 kV/cm was observed in 1T-TaS$_2$ at 20 K[66].

**APPENDIX D: The deviation of the 1/$I$ oscillations at high current**

As shown in Fig. 3(c) in the main text, there was a deviation of the 1/$I$ oscillations/jumps at high current from the strict periodicity, i.e., the voltage drops occurred at lower currents than the expected current values if following the linear relation of 1/$I_n$ vs. $n$, where $n$ is the oscillation index. We attribute such deviation at high current to be likely due to the heating effect, which might be suggested by the hysteresis observed at high current. As shown in Fig. 9(a), hysteresis was present when the current was swept above several tens $\mu$A on D2, but absent at low current. Furthermore, as depicted in Fig. 9(b), no hysteresis could be recognized when sweeping the current only up to 20 $\mu$A on D3.

In particular, D3 manifested nearly perfect periodic $1/I$ oscillations without any distinguishable deviations as shown in Fig. 5(a) in the main text. The deviation at high current was also observed on D1 as shown in Fig. 13. More subtle experimental and theoretical studies are required to elucidate the deviation.

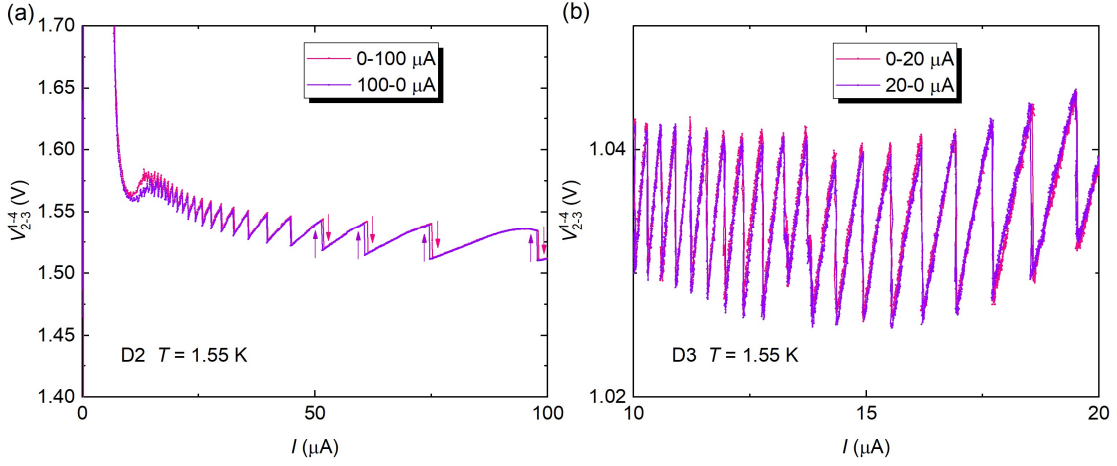

FIG. 9. (a) $V_{2-3}^{1-4} - I$ curve for device D2 at 1.55 K whose current was swept within a large loop, $0\ \mu\text{A} - 100\ \mu\text{A} - 0\ \mu\text{A}$. (b) $V_{2-3}^{1-4} - I$ curve for device D3 at 1.55 K whose current was swept within a small loop, $0\ \mu\text{A} - 20\ \mu\text{A} - 0\ \mu\text{A}$.

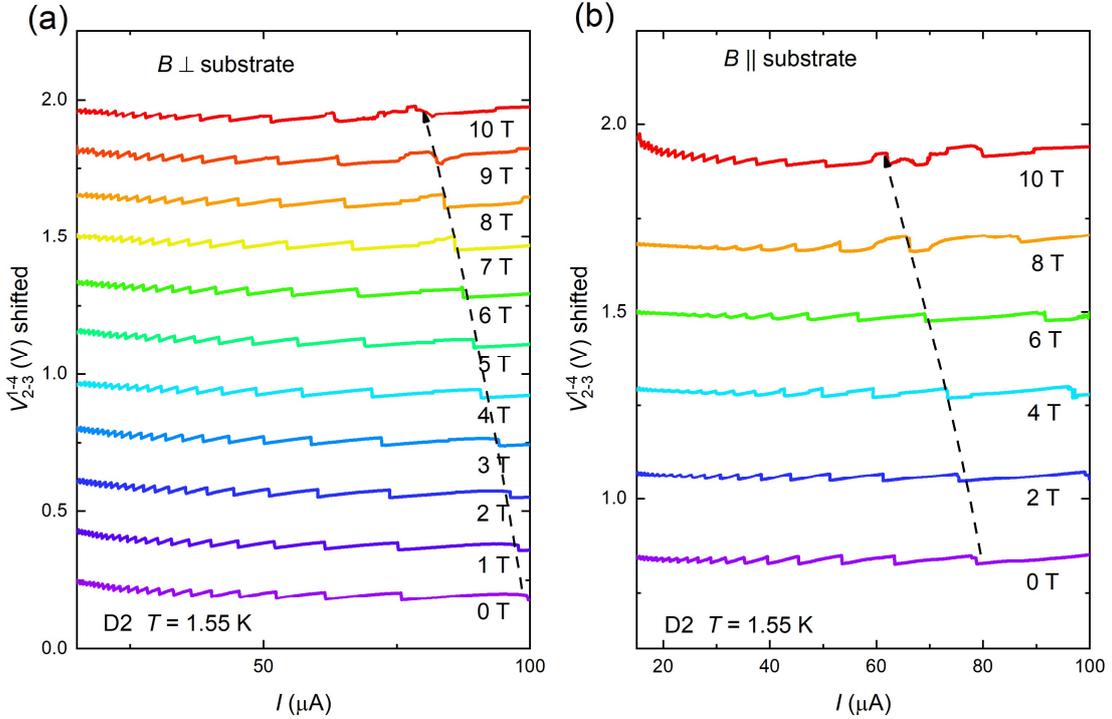

FIG. 10. (a), (b) Magnetic field evolution of the $V_{2-3}^{1-4} - I$ curves at 1.55 K for D2 when $B \perp$ and $B \parallel$ substrate, respectively.

## APPENDIX E: Shift of the current value for each voltage drop under a magnetic field

As is well known, the CDW state is a non-magnetic order and usually does not show any magnetic field-dependent behavior. Recently, the axionic CDW was proposed in $(TaSe_4)_2I$ and gave the magnetoelectric transport effects for the collinear electric and magnetic fields. In Fig. 4(c) for D3 in the main text, the magnetic field is parallel to the substrate and has a finite component along the electric field (current). Thus, we cannot exclude the contribution from the magnetoelectric effects for this single plot, and cannot determine if it is further related to the slight shift of the oscillations towards lower current. However, on D2, we observed such shift of the oscillations for both the perpendicular and parallel magnetic field, as shown in Fig. 10. As a result, the shift in magnetic field is likely irrelevant to the axionic CDW in $(TaSe_4)_2I$. We also note that the current value for each voltage drop is not shifted with increasing magnetic field on D6 and D7 using Ti/Nb electrodes, as shown in Figs. 11(c) and 11(f), which is in contrary to D1 and D2. The underlying details and mechanisms are still unclear for us and need further studies.

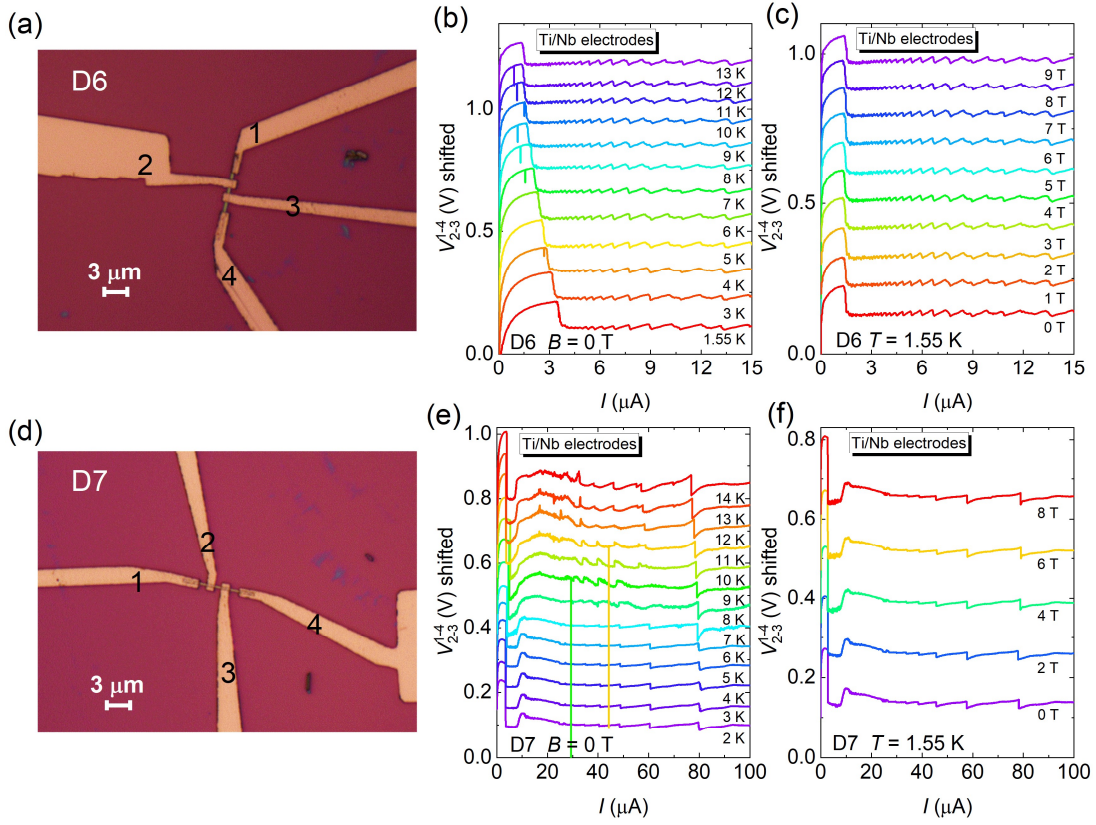

FIG. 11. (a), (d) Optical images for (TaSe$_4$)$_2$I nanowire devices D6 and D7 with the width of the nanowire of 450 nm and 350 nm, respectively. (b), (e) Temperature evolution of the $V_{2-3}^{1-4} - I$ curves at 0 T for D6 and D7, respectively. (c), (f) Magnetic field evolution of the $V_{2-3}^{1-4} - I$ curves at 1.55 K for D6 and D7, respectively.

**APPENDIX F: Niobium (Nb) superconducting electrodes on devices D6 and D7**

As we mentioned in the main text, the lower dissipation on current electrodes may be beneficial for observing the $1/I$ oscillations. However, it does not imply the electrodes must be a superconductor. We have fabricated two (TaSe$_4$)$_2$I nanowire devices D6 and D7 with the Ti/Nb (~5 nm/100 nm) electrodes as shown in Figs. 11(a) and 11(d). Ti/Nb is also a superconducting material with $T_c$ ~ 9 K while $H_{c2}$ < 2 T. Except for the difference of the critical magnetic field for Ti/NbTiN and Ti/Nb, there is a much smaller residual resistivity above the superconducting transition temperature for Ti/Nb than for Ti/NbTiN. At 1.55 K, both D6 and D7 show the nearly magnetic field-independent behavior as shown in Figs. 11(c) and 11(f), even if the superconductivity is killed above 2 T. Moreover, as displayed in Figs. 11(b) and 11(e), D6 still maintains the $1/I$ oscillations above $T_c$ of Ti/Nb with a gradual temperature smearing, while D7 shows the same trend but more smeary above $T_c$. These results demonstrate that electrodes of a normal metal can produce the 1/$I$ oscillations.

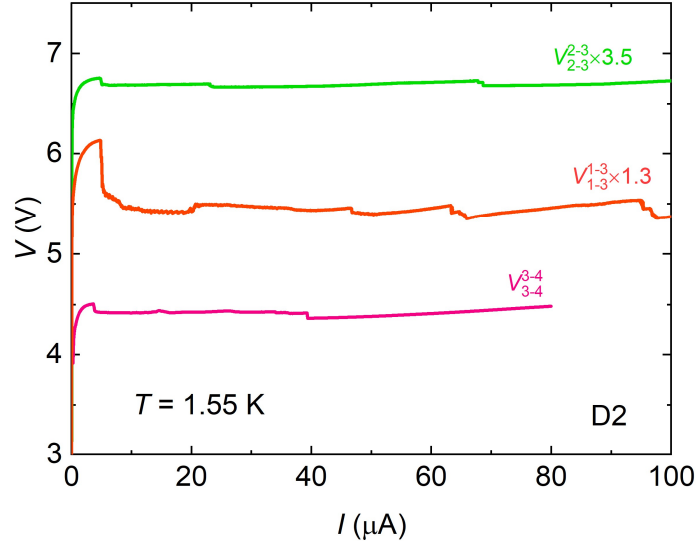

FIG. 12. *V-I* curves for three different pairs of electrodes on device D2 at 1.55 K, showing irregular oscillations.

**APPENDIX G: Intercept value *b* on the *n*-index axis**

For the de Haas–van Alphen (dHvA) or Shubnikov-de Haas (SdH) oscillations, according to the Lifshitz-Onsager relationship, the intercept value on the *n*-index axis depends on Berry phase, dimensionality of the Fermi surface and carrier type. However, we do not know the explicit physical meaning of the value *b* in our index analysis. The experimental results indicate that *b* could be influenced by the deviations (Appendix D) at high current. As shown in Fig. 13 for D1, the deviation from the periodic 1/$I$ oscillations (indicated as red dashed frame in the insets) is observed with increasing current for $V_{1-3}^{1-3} - I$, while the deviation is nearly neglectable for $V_{1-2}^{1-2} - I$ (Here, we define 1/$I_n$ as a minimum value for each oscillating unit, which is indicated by the red dots in Fig. 13(c). The deviation at high current probably generates a larger *b* for $V_{1-3}^{1-3} - I$ than for $V_{1-2}^{1-2} - I$. As mentioned in the Appendix D, it might be attributed to the heating effect resulted from the large current. Moreover, we also get a large *b* value (0.337) after analyzing the $V_{2-3}^{1-4} - I$ oscillations for D2 which showed deviations at high current, as plotted in Fig. 14, while *b* can be neglected for D3 presenting 1/$I$ oscillations at small current as shown in Figs. 5(a) and 5(b).

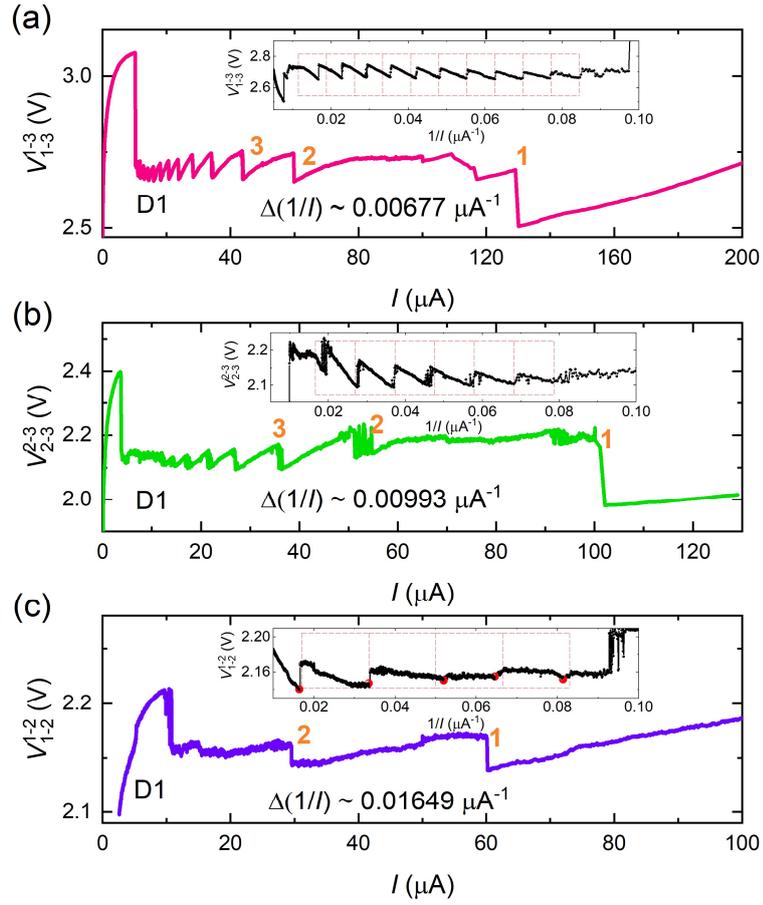

FIG. 13. (a), (b), (c) $V-I$ curves for electrodes $1-3$, $2-3$, and $1-2$ on D1, respectively. Insets are plotted as $V$ versus $1/I$.

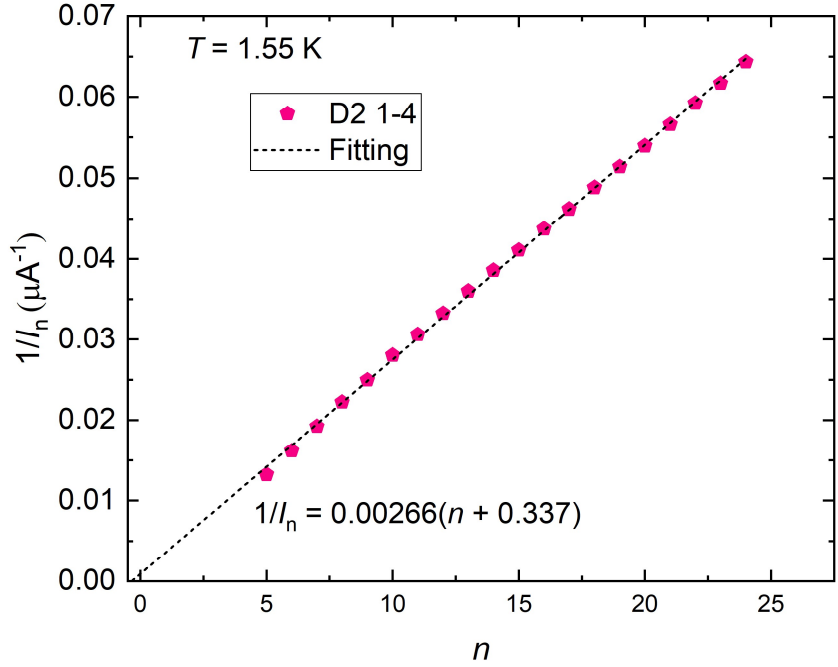

FIG. 14. Index $n$ dependence of the inverse current $1/I_n$ for D2. 1-4 denotes the current is applied between electrodes 1 and 4. The black dashed line is a linear fit.

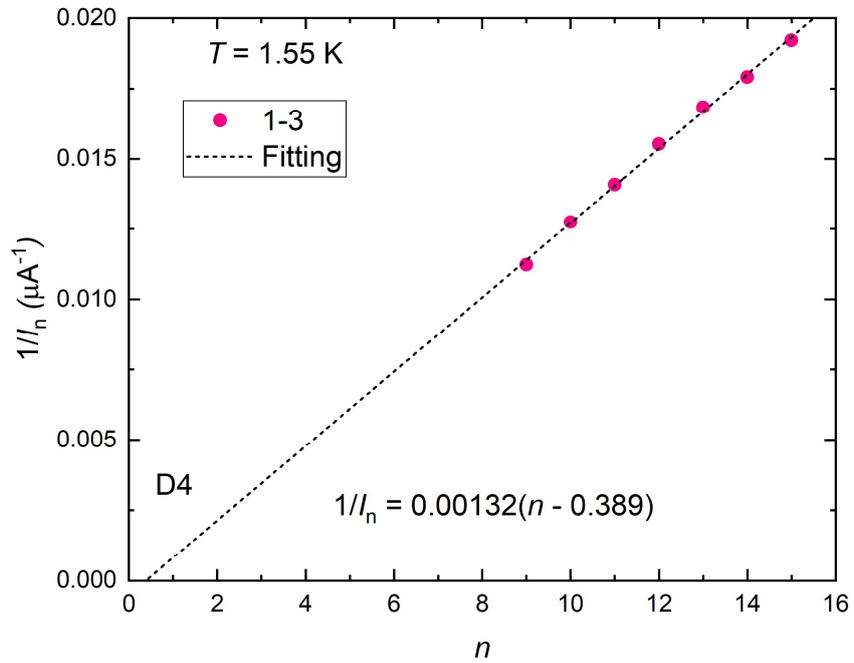

FIG. 15. Index $n$ dependence of inverse current $1/I_n$ for filtered $V$-$I$ curve on D4 at 1.55 K. 1-3 denotes the current is applied between electrodes 1 and 3. The black dashed line is a linear fit.

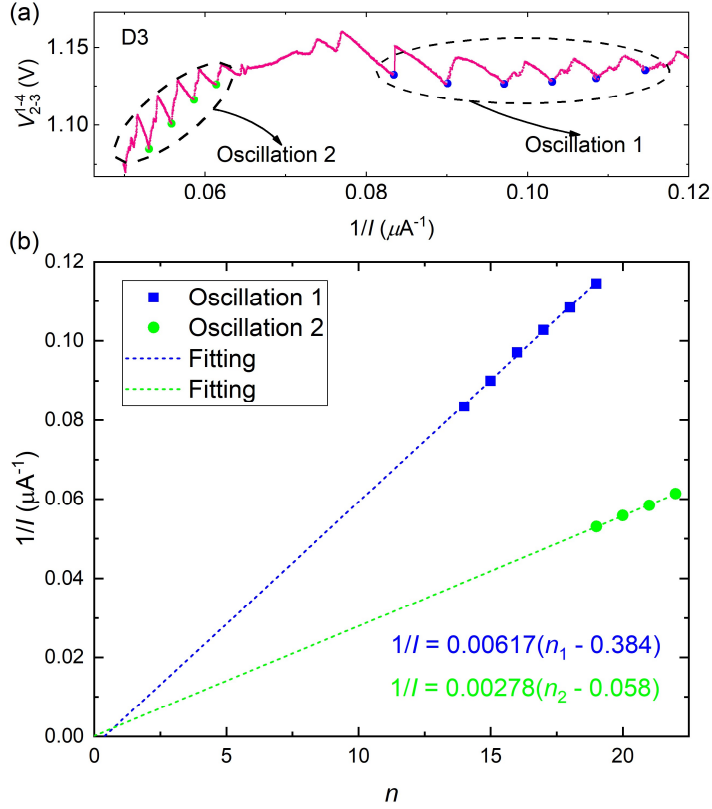

FIG. 16. (a) $V_{2-3}^{1-4}$ vs. $1/I$ for D3 at 15 K. (b) Index $n$ dependence of $1/I_n$ for oscillation region 1 and oscillation region 2, respectively. The blue and green dashed line represents a linear fit to the data.

**APPENDIX H: Analysis of re-entrance of oscillations at high temperatures**

We noticed the re-entrance of the oscillations at high temperatures in Fig. 4(a), and thermal smearing alone could not explain this feature. This is an interesting behavior which we did not find a clear interpretation yet. Nevertheless, we tried to analyze its period as shown in Fig. 16. Interestingly, we find that the period is different for the two segments of the curve. Furthermore, the first index of oscillation region 2 is larger than the last index of oscillation region 1, so that they belong to different groups of oscillations. Such re-entrance deserves further investigation.

**APPENDIX I: Excluding the artifacts from the measurement circuit**

Filters in the measurement circuit are constituted with capacitors and inductors to reduce the electrical noise. As mentioned in the main text, filters are helpful on the observation of the high-quality $1/I$ oscillations. In Fig. 17 we show that the $1/I$ oscillations could also be observed without the filters, although the quality of the oscillations is poor. Again, index $n$ dependence of the inverse current $1/I_n$ also exhibits a linear dependent behavior (Fig. 17(c)). Overall, the filters are beneficial for the high-quality $1/I$ oscillations, indicating the importance of quantum coherence in the device.

On the other hand, we noticed that the slope of the two $V$-$I$ curves is different in Fig. 6(a) for current values above ~30 μA. In our experiment, regardless of the filters, a slight change of the $V$-$I$ curves was frequently observed after a thermal cycle or after sweeping to a very large current (refer to Figs. 4(a) and 4(c) for example), indicating the device are sensitive to the external perturbations. Regarding to Fig. 6(a), there are a series of operations between the measurement of the two curves including sweeping to a large current and removing the filters in circuits, and thus the difference could be attributed to a slight change of the device.

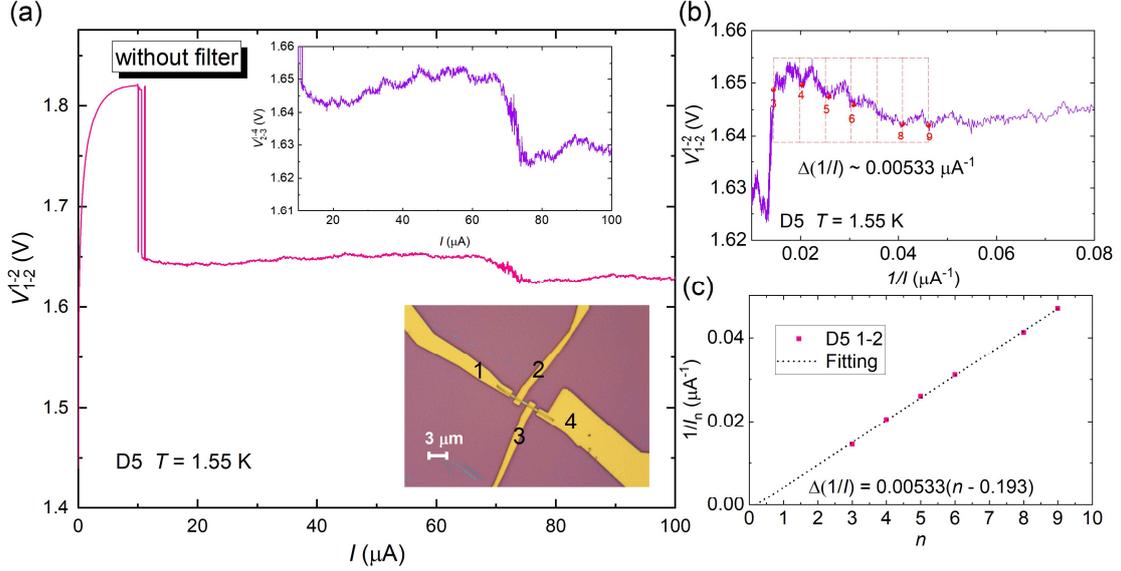

Fig. 17. (a) Unfiltered $V_{1-2}^{1-2} - I$ curve for (TaSe$_4$)$_2$I nanowire device D5 at 1.55 K. Inset is an optical image for D5 with the width of the nanowire around 400 nm. (b) $1/I$ dependence of $V_{1-2}^{1-2}$ at 1.55 K. (c) Index $n$ dependence of $1/I_n$ for D5. The black dashed line represents a linear fit to the data.

**APPENDIX J: Theoretical model**

We propose that a sliding CDW can be modeled by a time-dependent Hamiltonian explicitly. This time-dependent Hamiltonian is exactly solvable and gives rise to Floquet side-bands. Theoretically, the effects of quantum tunneling and quantum-electro ($1/I$-periodic) oscillation in such a driven system can be deduced in a mesoscopic device composed by a quasi-1D CDW insulator and metallic leads.

A static CDW is characterized by a periodic modulation of the charge density, $\rho(x) = \rho_0 + \rho_1 \cos(Qx + \phi)$, where $Q = 2k_F$ is the wave vector of the CDW. The simplest mean-field Hamiltonian for such a CDW state reads $H_{CDW} = \sum_k \epsilon_k c_k^+ c_k + \Delta e^{i\phi} c_{k+Q}^+ c_k + H.c.$ For a sliding CDW, the phase becomes time-dependent, $\phi(t) = \phi_0 - \omega_d t$. The diagonalization for the Floquet Hamiltonian $H_{CDW} - i\hbar \frac{\partial}{\partial t}$ leads to the quasi-energy dispersion $E_k = E_F + \left(m + \frac{1}{2}\right) \hbar \omega_d \pm \sqrt{\left(\left[\hbar v_F(k - k_F) + \frac{\hbar \omega_d}{2}\right]\right)^2 + \Delta^2}$ for quasiparticles. The integral Floquet spectral function is found to be $A_m(E) = A_0(E - m\hbar \omega_d)$, where $A_0(E) = \frac{1}{2} D\left(E + \frac{\hbar \omega_d}{2}\right) + \frac{1}{2} D\left(E - \frac{\hbar \omega_d}{2}\right)$ is the time-averaged density of states (DOS), and $D(E) = \frac{1}{\pi \hbar v_F} \frac{|E - E_F|}{\sqrt{(E - E_F)^2 - \Delta^2}} \theta(|E - E_F| - \Delta)$ is the DOS for the static CDW. A schematic illustration of the Floquet sidebands can be found in Fig. 6(d) in the main text.